\documentclass[reprint, superscriptaddress, twocolumn, amsmath, amssymb, aps, prb]{revtex4-2}
\usepackage{graphicx}
\usepackage{dcolumn}
\usepackage{bm}
\usepackage{placeins}
\usepackage{appendix}
\usepackage{upgreek}

\usepackage{verbatim}
\usepackage[usenames,dvipsnames]{xcolor}
 \usepackage{amsmath}
 \usepackage{amsfonts}
 \usepackage{amssymb}
 \usepackage[colorlinks=true,citecolor=blue,linkcolor=red]{hyperref}

 \usepackage{bbold}     
 \usepackage[makeroom]{cancel}  
 \usepackage{multirow}    
 \usepackage[normalem]{ulem}        
 \usepackage{array}
 \usepackage{pdfpages}
\makeatletter
 \AtBeginDocument{\let\LS@rot\@undefined}
 \makeatother

\begin{document}


\title{Quantized Andreev conductance in semiconductor nanowires}

\author{Yichun Gao}
\email{equal contribution}
\affiliation{State Key Laboratory of Low Dimensional Quantum Physics, Department of Physics, Tsinghua University, Beijing 100084, China}

\author{Wenyu Song}
\email{equal contribution}
\affiliation{State Key Laboratory of Low Dimensional Quantum Physics, Department of Physics, Tsinghua University, Beijing 100084, China}

\author{Yuhao Wang}
\email{equal contribution}
\affiliation{State Key Laboratory of Low Dimensional Quantum Physics, Department of Physics, Tsinghua University, Beijing 100084, China}

\author{Zuhan Geng}
\email{equal contribution}
\affiliation{State Key Laboratory of Low Dimensional Quantum Physics, Department of Physics, Tsinghua University, Beijing 100084, China}

\author{Zhan Cao}
\email{equal contribution}
\affiliation{Beijing Academy of Quantum Information Sciences, Beijing 100193, China}

\author{Zehao Yu}
\affiliation{State Key Laboratory of Low Dimensional Quantum Physics, Department of Physics, Tsinghua University, Beijing 100084, China}

\author{Shuai Yang}
\affiliation{State Key Laboratory of Low Dimensional Quantum Physics, Department of Physics, Tsinghua University, Beijing 100084, China}

\author{Jiaye Xu}
\affiliation{State Key Laboratory of Low Dimensional Quantum Physics, Department of Physics, Tsinghua University, Beijing 100084, China}

\author{Fangting Chen}
\affiliation{State Key Laboratory of Low Dimensional Quantum Physics, Department of Physics, Tsinghua University, Beijing 100084, China}

\author{Zonglin Li}
\affiliation{State Key Laboratory of Low Dimensional Quantum Physics, Department of Physics, Tsinghua University, Beijing 100084, China}

\author{Ruidong Li}
\affiliation{State Key Laboratory of Low Dimensional Quantum Physics, Department of Physics, Tsinghua University, Beijing 100084, China}

\author{Lining Yang}
\affiliation{State Key Laboratory of Low Dimensional Quantum Physics, Department of Physics, Tsinghua University, Beijing 100084, China}

\author{Zhaoyu Wang}
\affiliation{State Key Laboratory of Low Dimensional Quantum Physics, Department of Physics, Tsinghua University, Beijing 100084, China}

\author{Shan Zhang}
\affiliation{State Key Laboratory of Low Dimensional Quantum Physics, Department of Physics, Tsinghua University, Beijing 100084, China}

\author{Xiao Feng}
\affiliation{State Key Laboratory of Low Dimensional Quantum Physics, Department of Physics, Tsinghua University, Beijing 100084, China}
\affiliation{Beijing Academy of Quantum Information Sciences, Beijing 100193, China}
\affiliation{Frontier Science Center for Quantum Information, Beijing 100084, China}
\affiliation{Hefei National Laboratory, Hefei 230088, China}

\author{Tiantian Wang}
\affiliation{Beijing Academy of Quantum Information Sciences, Beijing 100193, China}
\affiliation{Hefei National Laboratory, Hefei 230088, China}

\author{Yunyi Zang}
\affiliation{Beijing Academy of Quantum Information Sciences, Beijing 100193, China}
\affiliation{Hefei National Laboratory, Hefei 230088, China}

\author{Lin Li}
\affiliation{Beijing Academy of Quantum Information Sciences, Beijing 100193, China}

\author{Dong E. Liu}
\affiliation{State Key Laboratory of Low Dimensional Quantum Physics, Department of Physics, Tsinghua University, Beijing 100084, China}
\affiliation{Beijing Academy of Quantum Information Sciences, Beijing 100193, China}
\affiliation{Frontier Science Center for Quantum Information, Beijing 100084, China}
\affiliation{Hefei National Laboratory, Hefei 230088, China}

\author{Runan Shang}
\affiliation{Beijing Academy of Quantum Information Sciences, Beijing 100193, China}
\affiliation{Hefei National Laboratory, Hefei 230088, China}

\author{Qi-Kun Xue}
\affiliation{State Key Laboratory of Low Dimensional Quantum Physics, Department of Physics, Tsinghua University, Beijing 100084, China}
\affiliation{Beijing Academy of Quantum Information Sciences, Beijing 100193, China}
\affiliation{Frontier Science Center for Quantum Information, Beijing 100084, China}
\affiliation{Hefei National Laboratory, Hefei 230088, China}
\affiliation{Southern University of Science and Technology, Shenzhen 518055, China}

\author{Ke He}
\email{kehe@tsinghua.edu.cn}
\affiliation{State Key Laboratory of Low Dimensional Quantum Physics, Department of Physics, Tsinghua University, Beijing 100084, China}
\affiliation{Beijing Academy of Quantum Information Sciences, Beijing 100193, China}
\affiliation{Frontier Science Center for Quantum Information, Beijing 100084, China}
\affiliation{Hefei National Laboratory, Hefei 230088, China}

\author{Hao Zhang}
\email{hzquantum@mail.tsinghua.edu.cn}
\affiliation{State Key Laboratory of Low Dimensional Quantum Physics, Department of Physics, Tsinghua University, Beijing 100084, China}
\affiliation{Beijing Academy of Quantum Information Sciences, Beijing 100193, China}
\affiliation{Frontier Science Center for Quantum Information, Beijing 100084, China}


\begin{abstract}

Clean one-dimensional electron systems can exhibit quantized conductance. The plateau conductance doubles if the transport is dominated by Andreev reflection. Here, we report quantized conductance observed in both Andreev and normal-state transports in PbTe-Pb and PbTe-In hybrid nanowires. The Andreev plateau is observed at $4e^2/h$, twice of the normal plateau value of $2e^2/h$. In comparison, Andreev conductance in the best-optimized III-V nanowires is non-quantized due to mode-mixing induced dips (a disorder effect), despite the quantization of normal-state transport. The negligible mode mixing in PbTe hybrids indicates an unprecedented low-disorder transport regime for nanowire devices, beneficial for Majorana researches.

\end{abstract}

\maketitle

Superconductor-semiconductor hybrid nanowires have been extensively studied for over a decade as promising candidates for the realization of Majorana zero modes \cite{Lutchyn2010, Oreg2010, Mourik, Deng2016, Gul2018, Song2022, WangZhaoyu, Delft_Kitaev, MS_2023, NextSteps, Prada2020}. Despite significant progress in material growth and device control \cite{Chang2015, Krogstrup2015, Gul2017, PanCPL}, current experiments remain limited by disorder. Disorder is also the key source of ongoing debates \cite{Patrick_Lee_disorder_2012, Prada2012, Loss2013ZBP, Liu2017, GoodBadUgly, DasSarma_estimate, DasSarma2021Disorder, Tudor2021Disorder}. Even the highest-quality nanowires, after a decade of optimization, still cannot meet the stringent low-disorder requirement for topological Majoranas \cite{DasSarma_Spectral}. Therefore, further reducing disorder is urgently needed to advance this field and to settle the debates.

A direct assessment of disorder in a nanowire device is through the observation of quantized conductance, a hallmark of ballistic (quasi) one-dimensional electron systems. In the early nanowire experiments around 2012, devices were highly disordered, and quantized conductance was absent at zero magnetic field. Later on, improvements in device fabrication have enabled the observation of zero-field quantized plateaus  in units of $2e^2/h$ in InSb and InAs devices configured in an N-NW-N geometry, where N stands for normal metal and NW for nanowire \cite{Kammhuber2016, InAs_Gooth, Silvano_ballistic}. 

This quantization is also observed in the normal-state conductance of N-NW-S hybrid devices \cite{Zhang2017Ballistic}, where S stands for superconductor. The normal-state conductance, $G_{\text{N}}$, is measured by applying a bias voltage significantly larger than the superconducting gap. The Andreev conductance ($G_{\text{S}}$), on the other hand, is measured for biases within the gap. In the tunneling regime, $G_{\text{S}}$ is suppressed relative to $G_{\text{N}}$, but it is enhanced in the open regime, i.e. the region of the $G_{\text{N}}$ plateau. This enhancement is due to Andreev reflection, wherein an injected electron is Andreev reflected as a hole, effectively doubling the charge transfer. Experimentally, this enhancement is rarely quantized due to significant fluctuations. In some regions on the $G_{\text{N}}$ plateau, the enhancement factor can reach the doubling limit, i.e.  $G_{\text{S}} \sim 4e^2/h$ \cite{Zhang2017Ballistic, Morten_doubling, Heedt_InSb_Al}, indicating a unity junction transmission ($T$) of Andreev process in those regions \cite{BTK}. The enhancement in $G_{\text{S}}$ is non-uniform and accompanied by large fluctuations, making it difficult to categorize as an Andreev plateau at $4e^2/h$. The reason for this is that $G_{\text{S}}$ scales with $T^2$, which is much more sensitive to disorder compared to $G_{\text{N}}$ which scales with $T$. When the second channel begins to be occupied, disorder-induced mode-mixing can suppress $G_{\text{S}}$, resulting in a dip feature. In contrast, $G_{\text{N}}$ is barely affected as the total transmission of the two channels remains unchanged. This disorder-induced dip, also revealed in numerical simulations \cite{Wimmer2011QPC, Zhang2017Ballistic}, is ubiquitous in the best-optimized III-V devices \cite{Zhang2017Ballistic, Morten_doubling, Heedt_InSb_Al}, preventing the formation of an Andreev plateau. While quantized Andreev conductance has long been predicted for clean N-NW-S devices \cite{1992_Beenakker, Wimmer2011QPC}, it has yet to be observed in nanowire experiments.         

\begin{figure*}[htb]
\includegraphics[width=0.75\textwidth]{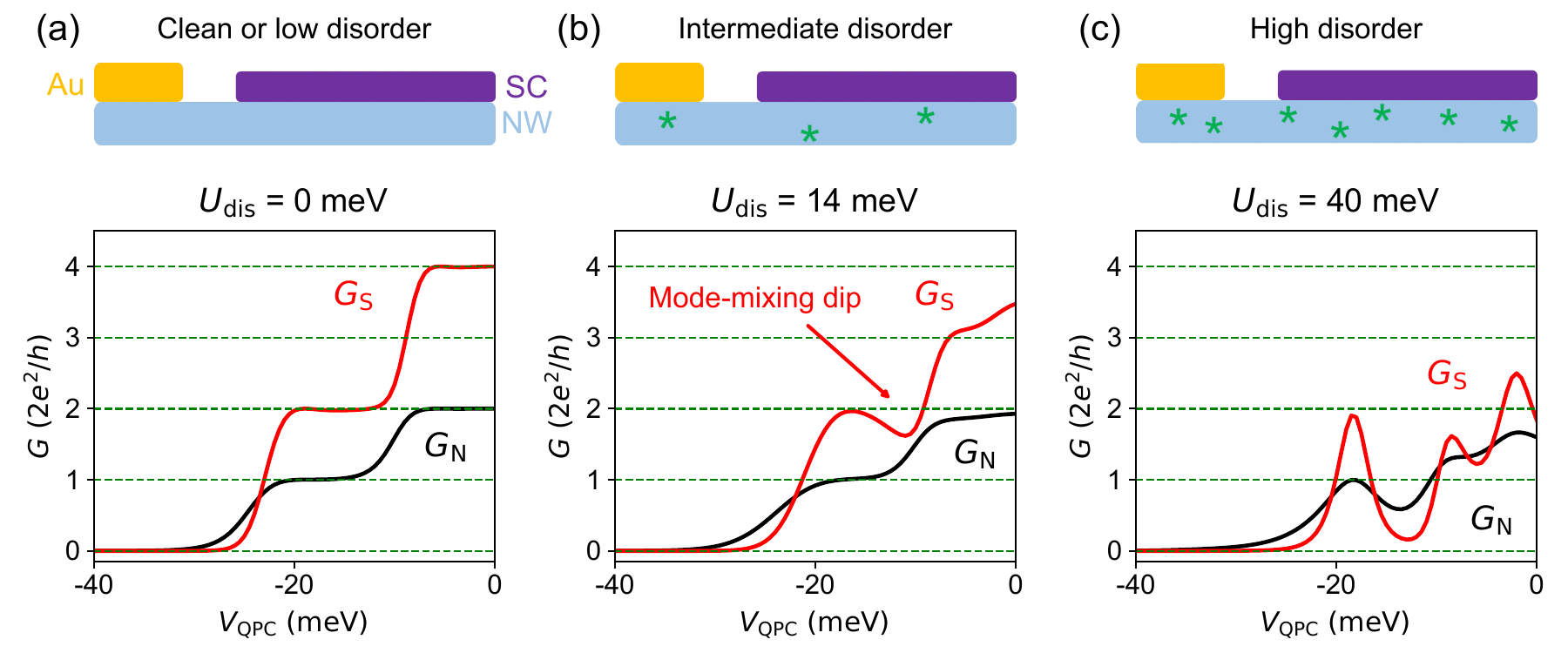}
\centering
\caption{Categorization of device disorder.  (a) Clean or low disorder regime. Upper, N-NW-S (yellow-blue-violet) device schematic. Lower panel, transport simulation: Both $G_{\text{S}}$ and $G_{\text{N}}$ exhibit quantized plateaus. $V_{\text{QPC}}$ refers the electrostatic potential of the barrier region, equivalent to a gate voltage. (b) Intermediate disorder regime. The green starts represent disorder for illustration purpose (not to scale). Lower panel, simulation reveals $G_{\text{N}}$ plateau, while the $G_{\text{S}}$ quantization is destroyed. The red arrow marks the characteristic dip induced by mode-mixing. (c) High disorder regime (illustrated as more green starts) with no quantization in either $G_{\text{S}}$ or $G_{\text{N}}$. Magnetic field is zero for the simulations. }
\label{fig2}
\end{figure*}

Here, we report the observation of quantized Andreev conductance in PbTe nanowires coupled to superconductors Pb and In. At zero magnetic field, $G_{\text{N}}$ is quantized at $2e^2/h$, exhibiting a diamond shape in the bias-gate 2D scan. The Andreev conductance is quantized at $4e^2/h$ without the mode-mixing-induced dip. At finite fields, superconductivity is suppressed, and the Andreev plateau transitions from $4e^2/h$ to $2e^2/h$. Our results suggest that transport in PbTe-Pb and PbTe-In nanowires achieves an unprecedented low-disorder regime compared to InAs and InSb nanowires. Given that PbTe has recently been proposed as a promising Majorana candidate, with rapid experimental advancements \cite{CaoZhanPbTe, Jiangyuying, Erik_PbTe_SAG, PbTe_AB, Fabrizio_PbTe, Zitong, Wenyu_QPC, Yichun, Yuhao, Ruidong, Vlad_PbTe, Wenyu_Disorder, PbTe_In, Yuhao_degeneracy, Yichun_SQUID}, these low-disorder hybrid nanowires may facilitate cleaner Majorana experiments in future research.

Figure 1 shows schematics of N-NW-S devices and numerical simulations of transport under varying disorder conditions. We categorize the disorder levels into three regimes: clean or low disorder (Fig. 1(a)), intermediate disorder (Fig. 1(b), and high disorder (Fig. 1(c)).  Green stars are sketched to represent different disorder levels for illustrative purpose (not to scale). In our theoretical model, disorder is introduced as on-site potential fluctuations within the nanowire junction, denoted as $\delta U_i \in [-U_{\text{dis}}, U_{\text{dis}}]$. $\delta U_i$ is the disorder potential at site $i$, randomly generated within the range $[-U_{\text{dis}}, U_{\text{dis}}]$. The disorder strength is defined by $U_{\text{dis}}$. Transport for each disorder distribution was calculated using the software package Kwant \cite{Kwant}, see Supplemental Materials (SM) for details. 

In the clean or low disorder regime, we set $U_{\text{dis}} = 0$. The calculated $G_{\text{N}}$ and $G_{\text{S}}$ are shown in the lower panel of Fig. 1(a), with $G_{\text{S}}$ obtained by setting the bias to zero. Both $G_{\text{N}}$ and $G_{\text{S}}$ exhibit quantized plateaus. The Andreev doubling, where $G_{\text{S}}$ = $2G_{\text{N}}$ on the plateau, is also revealed. 

In the intermediate disorder regime (Fig. 1(b)), we set $U_{\text{dis}}$ to 14 meV. The $G_{\text{N}}$ plateau at $2e^2/h$ remains, while the $G_{\text{S}}$ plateau is destroyed and accompanied by a characteristic dip (indicated by the red arrow) induced by mode mixing. This dip feature typically appears on the right side of the $G_{\text{N}}$ plateau where the second channel begins to be occupied. Assuming the normal-state transmissions of the first two channels are $T_1$ and $T_2$, then $G_{\text{N}} = (T_1+T_2)\times 2e^2/h$, and $G_{\text{S}} = [2T_1^2/(2-T_1)^2+2T_2^2/(2-T_2)^2] \times 2e^2/h$ \cite{BTK, 1992_Beenakker}. When only the first channel is occupied, $T_2$ = 0. As the second channel starts to be occupied (the right side of the $G_{\text{N}}$ plateau), $T_2$ increases and disorder-induced mode mixing becomes significant. This mixing barely affects the total transmission ($T_1+T_2$), thus the $G_{\text{N}}$ plateau remains stable. In contrast, $G_{\text{S}}$ can be significantly modified (suppressed) by the redistribution of the transmissions between $T_1$ and $T_2$, resulting the observed dip. The consistency between these simulations and previous observed transport phenomena suggests that the best-optimized III-V nanowires fall within this intermediate disorder regime  \cite{Zhang2017Ballistic, Morten_doubling, Heedt_InSb_Al}. 

Figure 1(c) shows the high disorder regime with $U_{\text{dis}}$ = 40 meV. Zero-field quantization is absent in both $G_{\text{S}}$ and $G_{\text{N}}$. The early nanowire experiments around 2012 fall within this category. For more simulations of different $U_{\text{dis}}$'s and distributions with different random on-site potentials, see Fig. S1 in SM.

\begin{figure}[htb]
\includegraphics[width=\columnwidth]{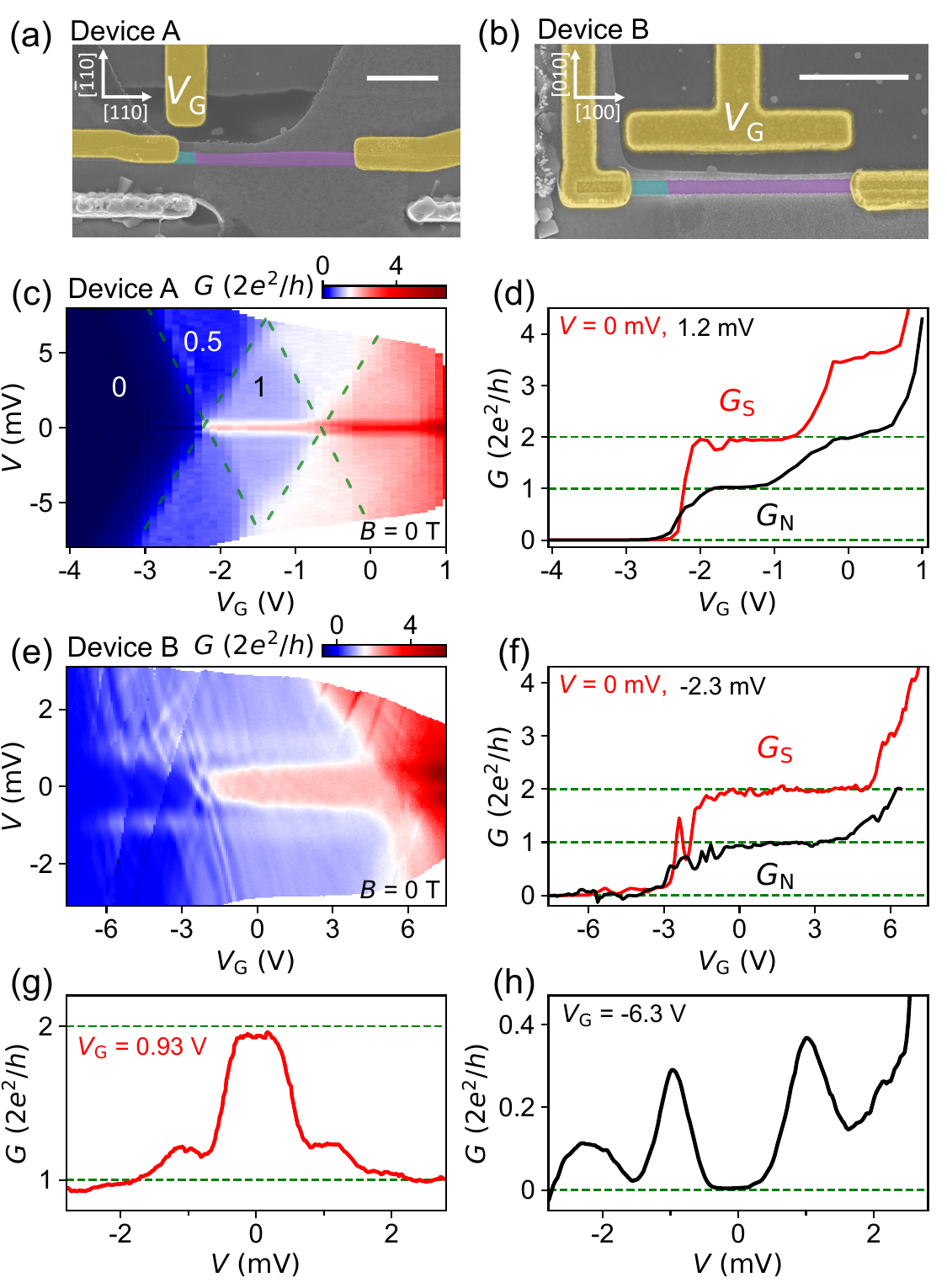}
\centering
\caption{Quantized Andreev conductance in PbTe-Pb nanowires at zero magnetic field. (a-b) False-colored SEM of two devices (A and B). Scale bars, 1 $\upmu$m.  (c) $G$ vs $V$ and $V_{\text{G}}$ for device A. (d) Extracted $G_{\text{S}}$ and $G_{\text{N}}$ from (c). $V$ = 0 mV and 1.2 mV, respectively. (e-f) Similar to (c-d) but for device B. (g-h) Line cuts from (e) in the open regime (on the $G_{\text{N}}$ plateau) and tunneling regime. $V_{\text{G}}$ = 0.93 V and -6.3 V, respectively. }
\label{fig2}
\end{figure}

Next, we present transport results of PbTe-Pb hybrid nanowires. Figures 2(a-b) are the scanning electron micrographs (SEMs) of two devices. The PbTe is false-colored blue, the superconductor Pb is violet, and contacts and gates (Ti/Au) are yellow. Growth details of these devices are provided in Ref. \cite{Wenyu_Disorder}. The fabrication process is identical to that in our previous works \cite{Wenyu_Disorder, PbTe_In, Yuhao_degeneracy, Yichun_SQUID}. The contact regions on the nanowire have been etched to remove the CdTe capping layer, ensuring ohmic contacts. Measurements were conducted in a dilution fridge at a base temperature of less than 50 mK using a standard two-terminal circuit. 

Figure 2(c) depicts the differential conductance $G \equiv dI/dV$ of device A as a function of bias voltage ($V$) and gate voltage ($V_{\text{G}}$).  $I$ is the current passing through the device. A series resistance, contributed by contact resistance ($R_c$) and the fridge filters, has been subtracted. For device A, $R_c$ = 620 $\Omega$. For a waterfall plot of Fig. 2(c), see Fig. S2 in SM. The green dashed lines delineate a diamond shape, indicating the quantization of $G_{\text{N}}$. The black curve in Fig. 2(d) is a line cut at $V$ = 1.2 mV (outside the gap), revealing the $G_{\text{N}}$ plateaus. The size of the diamond is $\sim$ 7.3 meV, which measures the subband spacing. The labeled numbers represent the plateau values in units of $2e^2/h$. Within the $2e^2/h$ diamond, $G$ near zero bias is significantly enhanced. The red curve in Fig. 2(d) shows the zero-bias line cut ($G_{\text{S}}$). $G_{\text{S}}$ exhibits a plateau feature near $4e^2/h$, doubling the value of the $G_{\text{N}}$ plateau. Notably, on the right side of the $G_{\text{N}}$ plateau where the second channel starts to be occupied, $G_{\text{S}}$ increases without dipping down, in sharp contrast to previous III-V experiments \cite{Zhang2017Ballistic, Morten_doubling, Heedt_InSb_Al}. The absence of the dip feature suggests negligible mode mixing, indicating that the device has achieved the low disorder regime depicted in Fig. 1(a). The small dip on the left side of the plateau is likely charge instability. This is unlikely to be mode mixing, as the second channel is far from being occupied. 

Figure 2(e) shows $G$ vs $V$ and $V_{\text{G}}$ for device B, where $R_c$ = 560 $\Omega$. The $G_{\text{S}}$ and $G_{\text{N}}$ line cuts are shown in Fig. 2(f), both exhibiting plateau features without the mode-mixing dip. Oscillations near pinch-off are observed in $G_{\text{S}}$ and $G_{\text{N}}$, which are likely Fabry-P\'{e}rot resonances. The visibility of Fabry-P\'{e}rot oscillations depends on the sharpness of the potential profiles within the wire. Even in clean wires, these oscillations are expected for a sharp potential profile (e.g. a square shape), and disappear for an adiabatically smoothed profile \cite{QPC_Smoothness}.  These oscillations occur near pinch-off and are thus not related to mode mixing, which occurs on the right side of the $G_{\text{N}}$ plateau.

Figure 2(g) shows bias spectroscopy on the Andreev plateau (a vertical line cut of Fig. 2(e)). The outside-gap conductance is $\sim 2e^2/h$, corresponding to the $G_{\text{N}}$ plateau. Near zero bias ($|V| <$ 0.4 meV), $G$ is enhanced/doubled to $4e^2/h$. Tuning the device into the tunneling regime (Fig. 2(h)), the subgap conductance is suppressed to zero, indicating a hard gap. The shoulders near $|V| \sim$ 1 mV in Fig. 2(g) are likely related to the broad coherence peaks revealed in Fig. 2(h). The quality of the gap is moderate, as the coherence peaks are not sharply defined. The gap of device A in Fig. 2(c) is worse, i.e., smaller and softer (see Fig. S2 in SM for line cuts). These results indicate that the quantization of $G_{\text{S}}$ and high-quality hard gap are not necessarily correlated.

We then investigate the magnetic field ($B$) dependence of the Andreev plateau. Figure 3(a) shows the $B$ scan of device B with $V$ fixed at 0 mV. $B$ is roughly parallel to the wire axis. The $4e^2/h$-Andreev plateau observed at zero field is suppressed and evolves into the $2e^2/h$-normal plateau at high fields, see Fig. 3(b) for line cuts. For a continuous evolution, Fig. 3(c) shows a vertical line cut on the plateau region of Fig. 3(a). $G_{\text{S}}$ drops sharply near 0.4 T, indicating a suppression of the induced superconductivity. The Fabry-P\'{e}rot resonances near pinch-off are barely affected by $B$. For line cuts at all fields, see Fig. S3 in SM.

\begin{figure}[htb]
\includegraphics[width=\columnwidth]{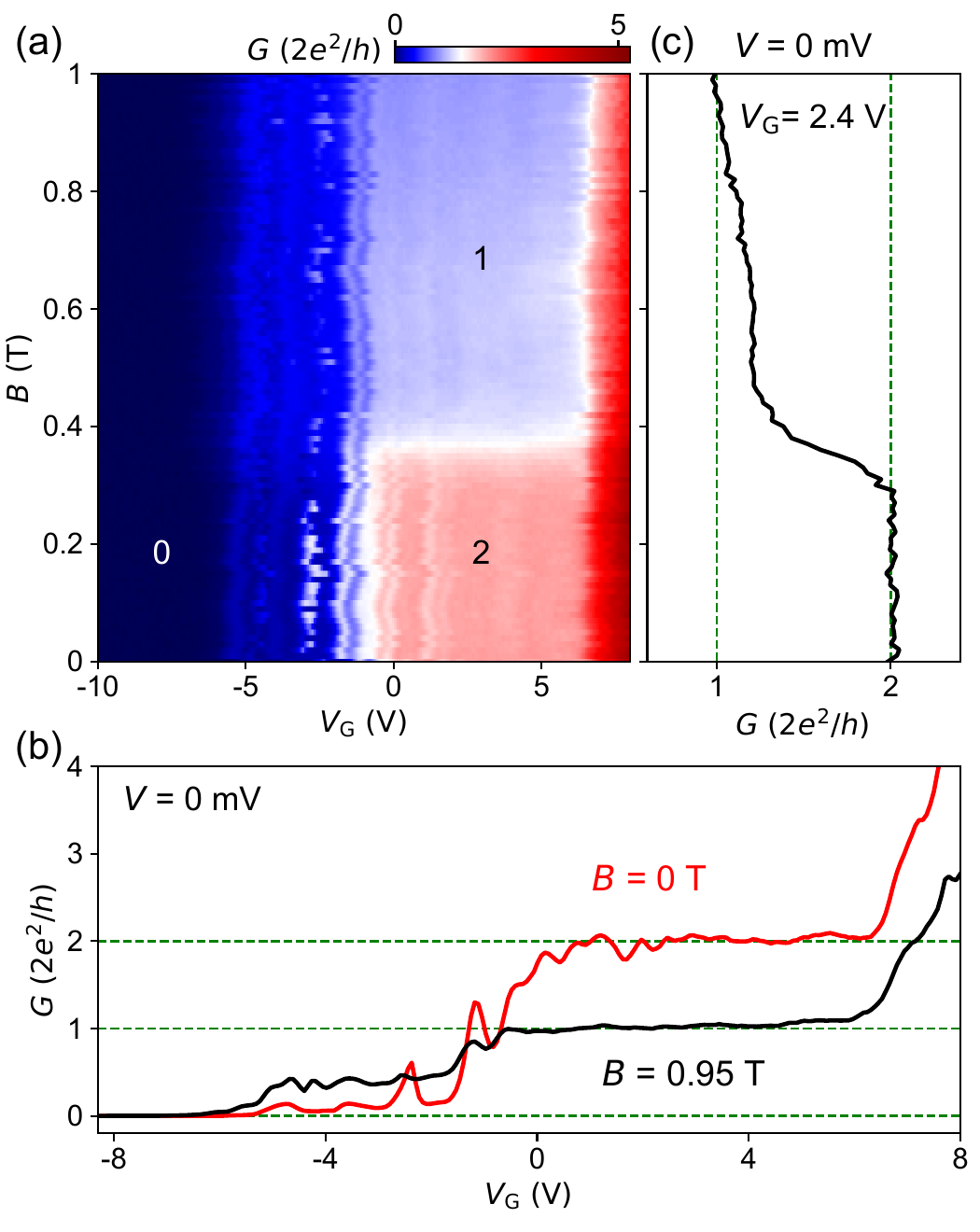}
\centering
\caption{$B$ dependence of the Andreev conductance in device B. (a) $G$ vs $B$ and $V_{\text{G}}$. $V$ = 0  mV. $B$ is roughly aligned with the wire axis. The numbers label the plateau values in units of $2e^2/h$. (b) Line cuts at $B$ = 0 T (red) and 0.95 T(black). (c) Line cut at $V_{\text{G}}$ = 2.4 V, illustrating the continuous evolution of the $G_{\text{S}}$ plateau. }
\label{fig2}
\end{figure}

The low critical field and broad coherence peaks are undesired in Majorana research and should be addressed in future optimizations. Although higher critical fields ($>$ 2 T) can be achieved for some devices \cite{Yichun}, the underlying mechanism of this device-to-device variation remains unclear. It might be related to the $g$-factor anisotropy in PbTe \cite{Fabrizio_PbTe, Yuhao}, as a $B$ parallel to the wire may induce a Zeeman splitting perpendicular to the wire direction \cite{CaoZhanPbTe}. Note that the superconducting gap is not protected if a Zeeman field is parallel to its spin-orbit direction \cite{Jouri2019}. Future optimization would require more systematic studies, such as exploring different wire geometries and symmetries. One possible solution for higher critical fields is to grow thinner wires to reduce the orbital effect \cite{Orbital_Anton, Loss_orbital}. Another avenue worth exploring is the implementation of a uniform barrier layer, such as Pb$_{0.99}$Eu$_{0.01}$Te, inserted in between PbTe and Pb. This spacing layer is nearly lattice-matched with PbTe, thereby reducing the disorder caused by the mismatch between the superconductor and semiconductor.  

\begin{figure}[tb]
\includegraphics[width=\columnwidth]{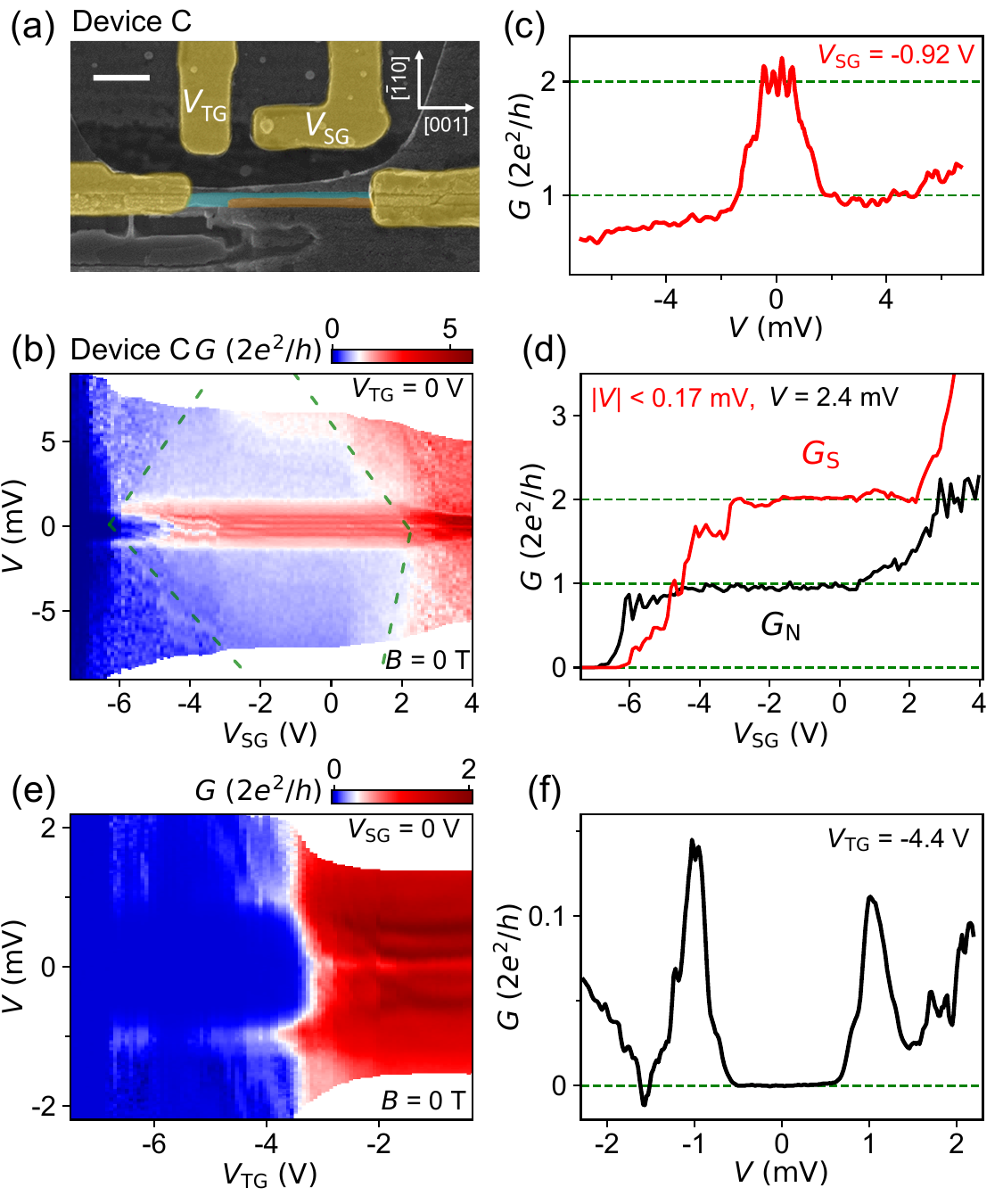}
\centering
\caption{Quantized Andreev conductance in a PbTe-In hybrid nanowire at $B$ = 0 T. (a) Device SEM. Scale bar, 500 nm. The indium film on PbTe is false-colored orange. (b) $G$ vs $V$ and $V_{\text{SG}}$. $V_{\text{TG}}$ = 0 V. (c) A vertical line cut from (b) on the plateau region. (d) Horizontal line cuts from (b), showing $G_{\text{S}}$ and $G_{\text{N}}$ ($V$ = 2.4 mV). $G_{\text{S}}$ is the average conductance within a bias window of $|V| <$ 0.17 mV to eliminate the oscillation effect in (c).  (e) Fine measurement in the tunneling regime. $V_{\text{SG}}$ = 0 V. (f) A line cut from (e) at $V_{\text{TG}}$ = -4.4 V. }
\label{fig2}
\end{figure}

The quantized Andreev conductance is observed not only in PbTe-Pb hybrids, but also in PbTe-In, as shown in Fig. 4.  Figure 4(a) presents the SEM of a PbTe-In nanowire. For the growth and transport characterization of PbTe-In hybrids, we refer to Ref. \cite{PbTe_In}.  Figure 4(b) shows the 2D conductance map of the device depicted in Fig. 4(a) at zero magnetic field. $R_c$ = 1.4 k$\Omega$. The green dashed lines delineate the diamond shape corresponding to the $G_{\text{N}}$ plateau, see the black curve in Fig. 4(d) for a line cut outside the gap. The red curve represents the conductance near zero bias (averaged over biases within 0.17 mV), revealing a quantized Andreev plateau around $4e^2/h$. The mode-mixing dip is also absent, indicating the low disorder regime. 

Figure 4(c) shows the bias spectroscopy on the plateau, corresponding to a line cut at $V_{\text{SG}}$ = -0.92 V. Within the superconducting gap ($|V| <$ 1 mV), the conductance is enhanced to $4e^2/h$. Sizable oscillations are superimposed on the enhanced plateau, also visible as horizontal lines in Fig. 4(b). The origin is not fully unclear but might be related to Fabry-P\'{e}rot resonances. Outside the gap ($|V| >$ 1 mV), the normal-state conductance exhibits a bias-dependent slope: it is slightly above $2e^2/h$ for positive $V$ and below $2e^2/h$ for negative $V$. Figure S4 in SM shows the waterfall plot of this $G_{\text{N}}$ asymmetry. Similar asymmetry has been observed in our previous devices \cite{Wenyu_QPC, Yuhao, Yichun}, potentially related to the cross-talk between gate and bias voltages. Figure 4(e) is a fine scan over a smaller bias range, focusing on the tunneling regime to reveal the superconducting gap. A hard gap is observed, see Fig. 4(f) for a line cut. The gap size, $\sim$ 1 meV, is consistent with our previous studies. The gap's critical field can reach 2 T, see Fig. S4 in SM.

In summary, we have observed quantized Andreev conductance in PbTe-Pb and PbTe-In hybrid nanowire devices. The absence of mode-mixing-induced dips suggests that our nanowires have reached the low disorder regime, which is unprecedented in comparison to previous studies on InAs and InSb nanowires. The observation indicates significant mitigation of disorder in PbTe nanowires, addressing a critical need in the field of Majorana research. These clean hybrid nanowires  may facilitate better signatures in future explorations of Majorana zero modes. Future optimization efforts could focus on improving the device yield and enhancing the critical magnetic field.

\textbf{Acknowledgment} This work is supported by National Natural Science Foundation of China (92065206, 12374158, and 12074039) and the Innovation Program for Quantum Science and Technology (2021ZD0302400). Raw data and processing codes within this paper are available at https://doi.org/10.5281/zenodo.11895952

\bibliography{mybibfile}

\newpage

\onecolumngrid

\newpage


\section*{Supplementary material}

# Supplemental Material for "Quantized Andreev conductance in semiconductor nanowires"

Yichun Gao,[1, *] Wenyu Song,[1, *] Yuhao Wang,[1, *] Zuhan Geng,[1, *] Zhan Cao,[2, *] Zehao Yu,[1] Shuai Yang,[1] Jiaye Xu,[1] Fangting Chen,[2] Zonglin Li,[1] Ruidong Li,[1] Lining Yang,[1] Zhaoyu Wang,[1] Shan Zhang,[1] Xiao Feng,[1, 2, 3, 4] Tiantian Wang,[2, 4] Yunyi Zang,[2, 4] Lin Li,[2] Dong E. Liu,[1, 2, 3, 4] Runan Shang,[2, 4] Qi-Kun Xue,[1, 2, 3, 4, 5] Ke He,[1, 2, 3, 4, †] and Hao Zhang[1, 2, 3, ‡]

[1] State Key Laboratory of Low Dimensional Quantum Physics,
Department of Physics, Tsinghua University, Beijing 100084, China
[2] Beijing Academy of Quantum Information Sciences, Beijing 100193, China
[3] Frontier Science Center for Quantum Information, Beijing 100084, China
[4] Hefei National Laboratory, Hefei 230088, China
[5] Southern University of Science and Technology, Shenzhen 518055, China

## Details of numerical simulations

We simulate the disorder effect using a two-dimensional effective Hamiltonian:

$$H = \left[ \frac{p_x^2 + p_y^2}{2m_{eff}} - \mu + V_{QPC}(x) + V_{dis}(x, y) \right] \tau_3 + \Delta(x) \tau_1, \tag{1}$$

with $m_{eff}$ the effective electron mass, $p_{x(y)}$ the momentum operator, $\mu$ the chemical potential, $V_{QPC}(x)$ the electrostatic potential forming the quantum point contact (QPC) region, $V_{dis}(x, y)$ the disorder-induced potential landscape, $\Delta$ the superconducting paring potential, and $\tau_{1(3)}$ the Pauli matrix acting on the electron-hole degree of freedom. $x$-axis refers to the longitudinal direction of the wire ($-\infty < x < \infty$), and $y$-axis refers to the transverse direction ($-W/2 < y < W/2$) ($W$ defines the wire width). The QPC, formed within the range of $-d/2 < x < d/2$, is contacted to a normal lead on the left side of the wire and to a superconducting lead [$\Delta(x) = \Delta_0 \theta(x - d/2)$] on the right side of the wire. $G_N$ is calculated by setting [$\Delta(x) = 0$]. We neglect spin-orbit coupling and Zeeman energy, and solely focus on the effect of disorder.

For simplicity, we employ a symmetric smooth QPC potential

$$V_{QPC}(x) = -\frac{eU_{QPC}}{2} \left( \tanh \frac{x + w/2}{\lambda} - \tanh \frac{x - w/2}{\lambda} \right), \tag{2}$$

where $U_{QPC}$, $w$, and $\lambda$ model the height, width, and smoothness, respectively. $V_{QPC}$ approaches $eU_{QPC}$ for $|x| < w/2$ and zero for $|x| > w/2 + \lambda$. We assume $d = 200$ nm, $w = 45$ nm, and $\lambda = 10$ nm. $U_{QPC}$ varies from -40 meV to 0 meV to simulate the effect of a gate voltage, i.e. the $x$-axis of the simulation plots in Fig. 1 and Fig. S1. In the QPC region, the spatially varying disorder potential $V_{dis}(x, y)$ is modeled by random numbers obeying the uniform distribution within the range $[-U_{dis}, U_{dis}]$. Note that the model is not completely identical to the actual device, as the nanowire in the actual device has a finite length and is contacted by source/drain metal leads (while in the model it is infinitely long without metal leads). To account for this difference, the disorder potential is introduced only in the QPC region.

We focus on the zero-bias conductance at zero temperature, which can be calculated by the scattering matrix formalism [1]:

$$G = \frac{e^2}{h} \text{Tr} \left[ I - |S_{ee}(0)|^2 + |S_{he}(0)|^2 \right]. \tag{3}$$

The scattering matrix $S_{ee}(0)$ and $S_{he}(0)$ are associated with the normal and Andreev reflections, respectively, of electrons at the Fermi level (i.e., $E = 0$). They can be readily obtained with the Kwant package [2] by discretizing the Hamiltonian (1) on a square lattice with a spacing of $a$. In all simulations, we adopt the parameters $m_{eff} = 0.04$ $m_e$ ($m_e$ is the free electron mass), $W = 40$ nm, $\mu = 30$ meV, $\Delta_0 = 0.4$ meV, and $a = 2$ nm.

—————
* equal contribution
† kehe@tsinghua.edu.cn
‡ hzquantum@mail.tsinghua.edu.cn

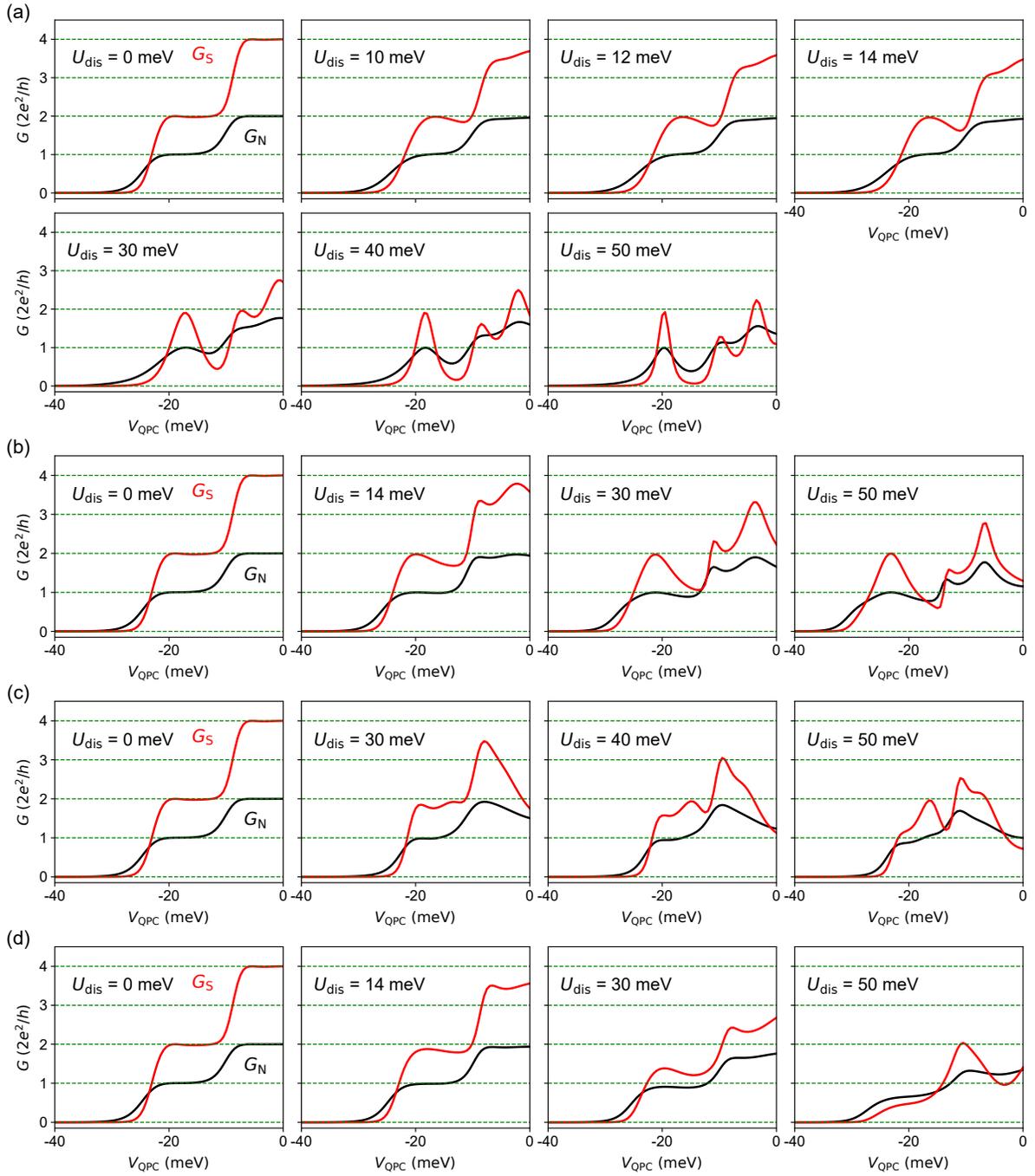

FIG. S1. Additional simulations. (a) Disorder distribution same with that in Fig. 1. We first generated $\delta U_i \in [-1 \text{ meV}, 1 \text{ meV}]$, and then scaled $\delta U_i$ by a factor of 0, 10, 12, 14, 30, 40, and 50 to obtain the transport behaviors shown in the 5 panels, respectively. The 0, 14, and 40 meV cases have been shown in Fig. 1. (b-d) Another three random distributions of $\delta U_i$, showing similar behavior. The Andreev plateau is in general more sensitive to disorder than the normal-state plateau.

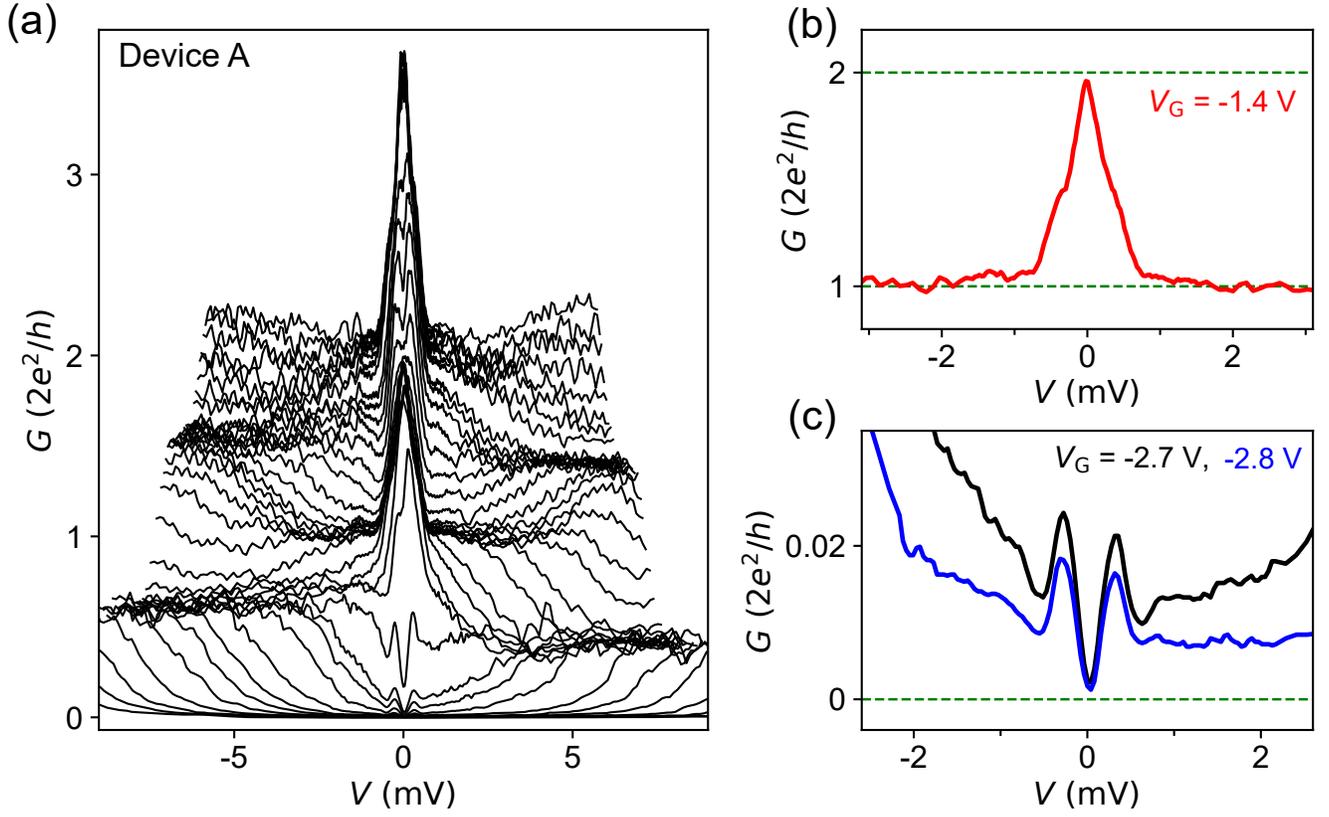

FIG. S2. (a) Waterfall plot of Fig. 2(c). The gate voltage range is from -3.5 V to 0.4 V for clarity. Plateaus are revealed as clusters of line cuts. At low biases (outside the gap), the $2e^2/h$ plateau is observable, while half plateaus (near $0.5 \times 2e^2/h$ and $1.5 \times 2e^2/h$) can be revealed at higher biases. (b) Bias scan on the plateau, corresponding to a line cut at $V_G = -1.4$ V. The outside gap conductance reaches the $2e^2/h$ normal plateau, the zero-bias conductance is close to $4e^2/h$. (c) Bias can in the tunneling regimes, resolving a soft gap.

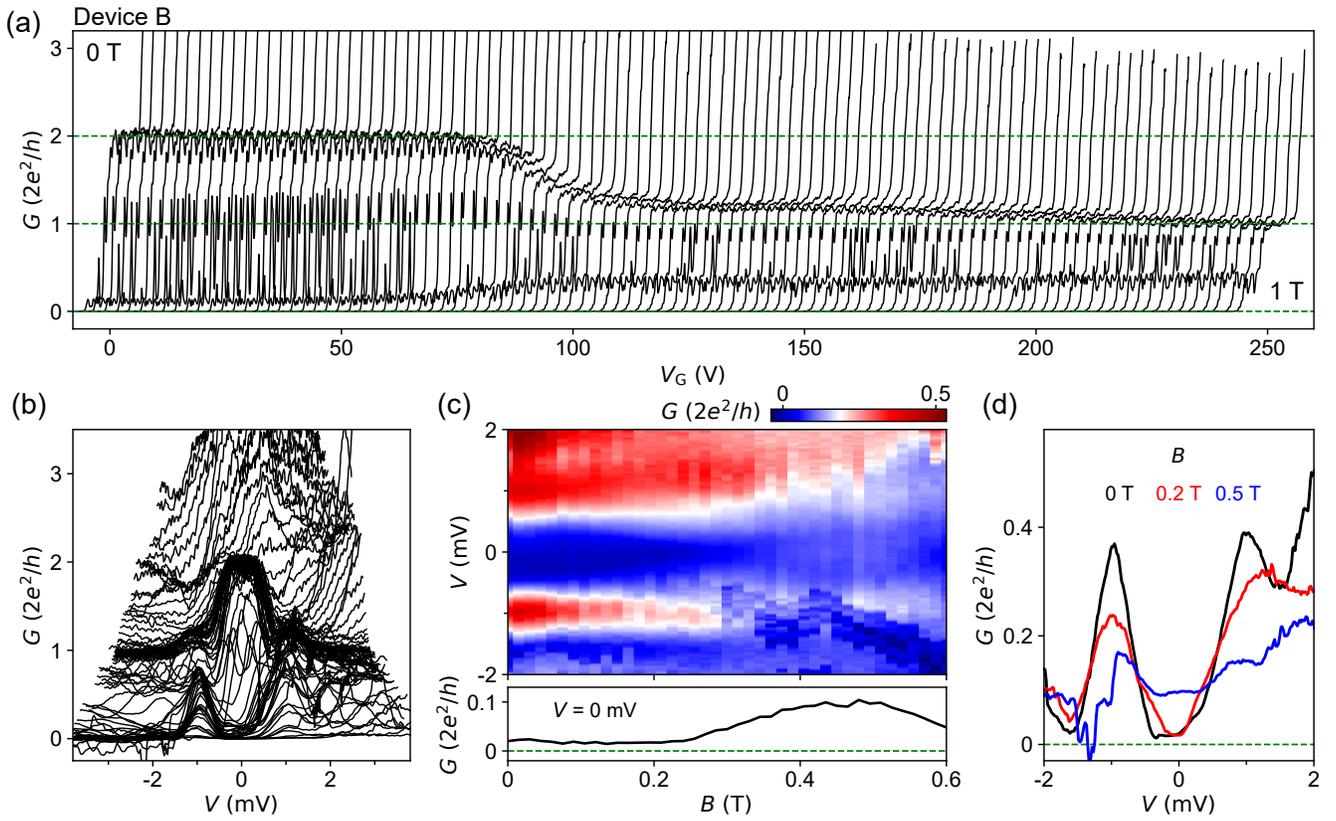

FIG. S3. (a) All line cuts of Fig. 3(a), illustrating the gradual evolution of the Andreev plateau in $B$. Horizontal offset between neighboring curves is 2.5 V for clarity. (b) Waterfall plot of Fig. 2(e), showing the $4e^2/h$ Andreev plateau and $2e^2/h$ normal plateau as clusters of line cuts. (c) $B$ dependence of the gap. $B$ is parallel to the wire axis. Lower panel, zero-bias line cut. The gap closes around 0.4 T, consistent with the transition field of the Andreev plateau shown in Fig. 3(a). (d) Line cuts from (c) at different fields.

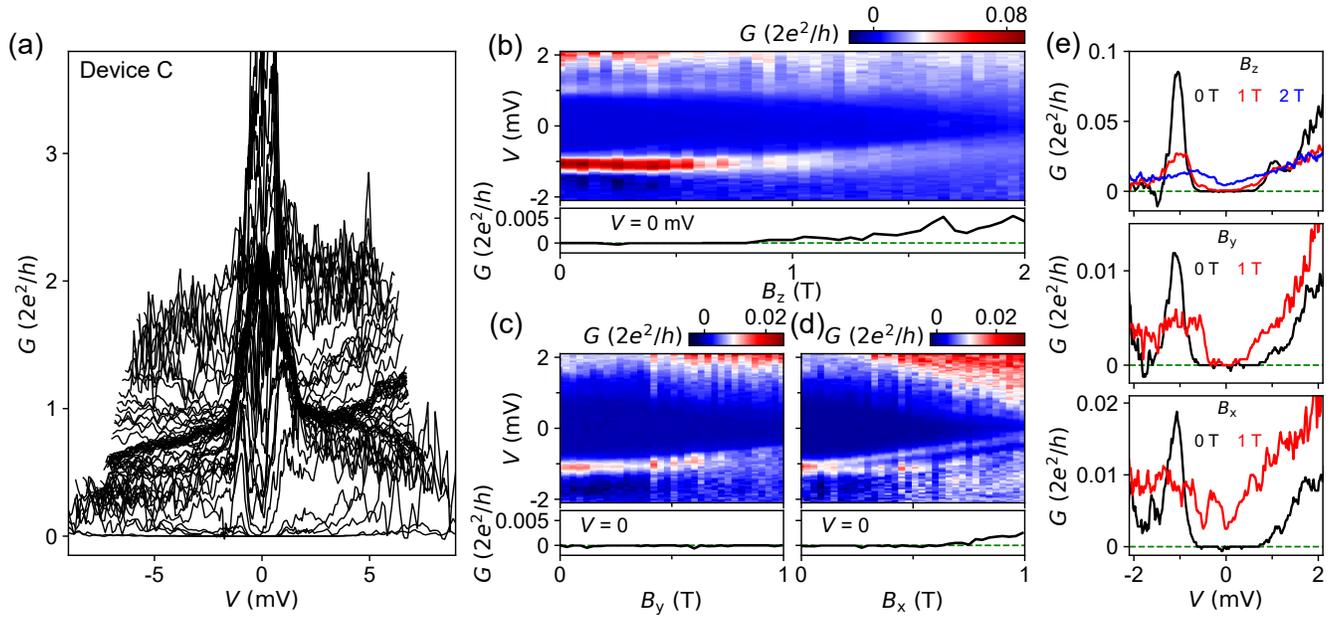

FIG. S4. (a) Waterfall plot of Fig. 4(b). (b-d) $B$ dependence of the gap along three axis. $B_z$ is parallel to the wire, $B_x$ is perpendicular to the wire and in-plane, $B_y$ is out of plane. Lower panels are zero-bias line cuts. (e) Line cuts at different fields. The critical field along the $z$ axis is near 2 T.



\end{document}